\documentclass[12pt]{article}
\usepackage{epsfig}
\def\be{\begin{equation}}
\def\ee{\end{equation}}
\def\bea{\begin{eqnarray}}
\def\eea{\end{eqnarray}}
\def\beq{\begin{equation}}
\def\eeq{\end{equation}}
\def\bq{\begin{quote}}
\def\eq{\end{quote}}
\def\gappeq{\mathrel{\rlap {\raise.5ex\hbox{$>$}} {\lower.5ex\hbox{$\sim$}}}}
\def\lappeq{\mathrel{\rlap{\raise.5ex\hbox{$<$}} {\lower.5ex\hbox{$\sim$}}}}
\def\PR{{\it Phys.~Rev.~}}

\def\NP{{\it Nucl.~Phys.~}}

\def\PL{{\it Phys.~Lett.~}}

\def\SJNP{{\it Sov.~J.~Nucl.~Phys.~}}

\def\JHEP{{\it J.~High~En.~Phys.~}}
\def\vol#1{{\bf #1}}
\def\vyp#1#2#3{\vol{#1} (#2) #3}

\def\epm#1#2{\hbox{${\lower1pt\hbox{$\scriptstyle +#1$}}
\atop {\raise1pt\hbox{$\scriptstyle -#2$}}$}}

\def\gsim{\mathrel{\rlap{\lower4pt\hbox{\hskip1pt$\sim$}}
    \raise1pt\hbox{$>$}}}         

\def\ie{{\it i.e.}}

\def\etal{{\it et al.}}

\def\toinf#1{\mathrel{\mathop{\sim}\limits_{\scriptscriptstyle
{#1\rightarrow\infty }}}}

\def\frac#1#2{{{#1}\over {#2}}}
\def\half{\hbox{${1\over 2}$}}

\def\smallfrac#1#2{\hbox{${{#1}\over {#2}}$}}

\def\as{\alpha_s}

\def\MS{\hbox{$\overline{\rm MS}$}}

\def\bq{\bar{q}}
\catcode`@=11 
\def\slash#1{\mathord{\mathpalette\c@ncel#1}}
 \def\c@ncel#1#2{\ooalign{$\hfil#1\mkern1mu/\hfil$\crcr$#1#2$}}
\def\lsim{\mathrel{\mathpalette\@versim<}}
\def\gsim{\mathrel{\mathpalette\@versim>}}
 \def\@versim#1#2{\lower0.2ex\vbox{\baselineskip\z@skip\lineskip\z@skip
       \lineskiplimit\z@\ialign{$\m@th#1\hfil##$\crcr#2\crcr\sim\crcr}}}
\catcode`@=12 

\begin{document} 
\pagestyle{empty} 
\begin{flushright} 
{\tt hep-ph/9911273}\\{CERN-TH/99-317}\\
{RM3-TH/99-11}\\ {Edinburgh 99/18}\end{flushright} \vspace*{5mm}
\begin{center} {\large\bf Resummation of Singlet Parton Evolution
at Small x}\\
\vspace*{0.8cm} {\bf Guido Altarelli,$^{a,\,b}$ 
Richard D. Ball$^{d,\;}$}\footnote[1]{Royal Society University Research
Fellow} 
and 
{\bf Stefano Forte$^{c,\;}$}\footnote[2]{On leave from INFN, Sezione di
Torino, Italy} \\
\vspace{0.6cm}
{}$^a$Theory Division, CERN\\ CH--1211 Geneva 23, Switzerland \\
\vspace{0.4cm}
{}$^b$Universit\`a di Roma Tre \\
{\it and}\\
{}$^c$INFN, Sezione di Roma Tre\\
Via della Vasca Navale 84, I-00146 Rome, Italy\\
\vspace{0.4cm}
{}$^d$Department of Physics and Astronomy, University of Edinburgh,\\
Mayfield Road, Edinburgh EH9 3JZ, Scotland\\

\vspace*{0.9cm}
{\bf Abstract}
\end{center}
\noindent
We propose an improvement of the splitting functions at small $x$ which
overcomes the apparent problems encountered by the BFKL approach. We
obtain a stable expansion for the $x$--evolution function $\chi(M)$
near $M=0$ by including in it a sequence of terms derived from the
one-- and two--loop anomalous dimension $\gamma$. The requirement of momentum
conservation is always satisfied.
The residual ambiguity on the splitting functions is effectively parameterized
in terms of the value of $\lambda$, which fixes the small $x$
asymptotic behaviour $x^{-\lambda}$ of the singlet parton distributions. We
derive from this improved evolution function an expansion of the
splitting function which leads to good apparent convergence, and to
a description of scaling violations valid both at large
and small $x$.
\\
\vspace*{0.5cm}

\vfill
\noindent

\begin{flushleft} CERN-TH/99-317 \\ October 1999 \end{flushleft} 
\eject 

\setcounter{page}{1} \pagestyle{plain}

\noindent

{\bf 1.}~~~The theory of scaling violations in deep inelastic scattering 
is one of the most
solid consequences of asymptotic freedom and provides a set of fundamental
tests of QCD. At large $Q^2$ and not too small but fixed $x$ the QCD
evolution equations for parton densities~\cite{glap} provide the
basic framework for the description of scaling violations. The
complete splitting functions have been computed in perturbation theory at order
$\alpha_s$ (LO  approximation) and $\alpha_s^2$ (NLO)~\cite{nlo}. For the first
few moments the anomalous dimensions  at order $\alpha_s^3$  are also
known~\cite{nnlo}. 

At sufficiently small $x$ the approximation of the splitting
functions based on  the first few terms in the expansion 
in powers of $\alpha_s$ is not in general a good approximation. 
If not for other reasons, as soon as $x$ is small enough that 
$\alpha_s \xi\sim 1$, with $\xi=\log{1/x}$, all terms of order
$\alpha_s (\alpha_s \xi)^n$ and $\alpha_s^2 (\alpha_s \xi)^n$ which
are present~\cite{kis} in the splitting functions must
be considered in order to achieve an accuracy up to terms of order
$\alpha_s^3$. In terms of the anomalous dimension $\gamma(N,\alpha_s)$,
defined as the $N$--th 
Mellin moment of the singlet splitting function (actually the
eigenvector with largest eigenvalue), these terms correspond to
sequences of the form $(\alpha_s/N)^n$ or $\alpha_s(\alpha_s/N)^n$.  
In most of the kinematic region of HERA~\cite{klein} the condition $\alpha_s
\xi\sim 1$ is indeed true. Hence,  in principle one could expect to
see in the data indications of important
corrections to the approximation~\cite{DGPTWZ,das} 
of splitting functions computed 
only  up to order $\alpha_s^2$ and the corresponding
small $x$ behaviour. In reality this
appears not to be the case: the data can be fitted quite well by the
evolution equations in the NLO approximation~\cite{das,botje}. 
Of course it may be that
some corrections exist but they are hidden in a
redefinition of the gluon, which is the dominant parton density at
small $x$. While the data do not support the presence of large
corrections in the HERA kinematic region~\cite{bfklfits} the evaluation of the 
higher order corrections at small $x$ to the singlet splitting function from
the BFKL theory~\cite{bfkl,jar,ktfac} appears to fail. 
The results of the recent calculation~\cite{fl,cc,dd}
of the NLO term $\chi_1$ of the BFKL function
$\chi=\alpha_s\chi_0+\alpha_s^2\chi_1...$ show that the expansion is very badly
behaved, with the non leading term completely overthrowing the main
features of the leading term. Taken at face value, these results appear 
to hint at very large corrections to the singlet splitting function at
small $x$ in the region explored by HERA~\cite{flph}.

In this article we address this problem and propose a 
procedure to construct a meaningful improvement of the singlet
splitting function at small $x$, using the information from the
BFKL function. We
start by defining an alternative expansion for the BFKL function
$\chi(M)$ which, unlike the usual expansion, is well behaved and stable
when going from LO to NLO, at least for values far from $M=1$.
This is obtained by adding suitable
sequences of terms of the form $(\alpha_s/M)^n$ or $\alpha_s(\alpha_s/M)^n$ to
$\as\chi_0$ or $\as^2\chi_1$ respectively. The coefficients are determined by
the known form of the singlet anomalous dimension at one and two loops. 
This amounts to a resummation~\cite{salam} of $(\alpha_s\log{Q^2/\mu^2})^n$
terms in the inverse $M$-Mellin transform space.
This way of improving $\chi$ is completely analogous to the usual way  
of improving~$\gamma$~\cite{summ}. One important point,
which is naturally reproduced with good accuracy by the above procedure,
is the observation that the value of $\chi(M)$ at $M=0$ is fixed by momentum
conservation to be $\chi(0)=1$. This observation plays a
crucial role in formulating the novel expansion 
and explains why the normal BFKL expansion is so unstable
near $M=0$, with $\chi_0\sim 1/M$, $\chi_1\sim -1/M^2$ and so on. 
This rather model--independent step is already sufficient to show that
no catastrophic deviations from the NLO approximation of the
evolution equations are to be expected.
The next step is to use this novel expansion of  $\chi$ to determine
small $x$ resummation corrections  to add to the LO and NLO anomalous
dimensions $\gamma$. Defining $\lambda$ as the minimum value of $\chi$,
$\chi(M_{min})=\lambda$, and
using the results of ref.~\cite{sxap}, a meaningful expansion for
the improved anomalous dimension is written down in terms of $\chi_0$,
$\chi_1$, and $\lambda$. The large negative
correction to $\lambda_0/\as=\chi_0(1/2)$ induced by $\as\chi_1$, 
that is formally of order $\as$ but actually is of order 
one for the relevant values of $\alpha_s$, suggests that $\lambda$ should be
reinterpreted as a nonperturbative parameter. 
We conclude by showing that the very good agreement of the data 
with the NLO evolution equation can be obtained by choosing a 
small value of $\lambda$,  compatible with zero.

\bigskip\noindent
{\bf 2.}~~~We consider the singlet parton density
\beq
G(\xi,t)=x[g(x,Q^2)+k_q\otimes q(x,Q^2)],\label{Gdef}
\eeq
where $\xi=\log{1/x}$, $t=\log{Q^2/\mu^2}$, 
$g(x,Q^2)$ and $q(x,Q^2)$ are the gluon and singlet quark parton densities,
respectively, and $k_q$ is such that, for each moment
\beq
G(N,t)=\int^{\infty}_{0}\! d\xi\, e^{-N\xi}~G(\xi,t),\label{Nmom}
\eeq
the associated anomalous dimension $\gamma(N,\as(t))$ corresponds to the
largest eigenvalue in the singlet sector. At large $t$ and fixed $\xi$
the evolution equation in $N$-moment space is then
\beq
\frac{d}{dt}G(N,t)=\gamma(N,\as(t))~G(N,t),\label{tevol}
\eeq
where $\as(t)$ is 
the running coupling. The anomalous dimension is completely known 
at one and two loop level:
\beq
\gamma(N,\as)=\as\gamma_0(N)+\as^2\gamma_1(N)+
\dots\>.\label{gamexp}
\eeq
As $\gamma(N,\as)$ is, for each $N$, the largest eigenvalue 
in the singlet sector, momentum conservation order by order 
in $\as$ implies that
\beq
\gamma(1,\as)=\gamma_{0}(1)=\gamma_{1}(1)=...=0.\label{mcons}
\eeq

Similarly, at large $\xi$ and fixed $Q^2$, the 
following evolution equation for $M$ moments is valid:
\beq
\frac{d}{d\xi}G(\xi,M)=\chi(M,\as)~G(\xi,M),\label{xevol}
\eeq
where
\beq
G(\xi,M)=\int^{\infty}_{-\infty}\! dt\, e^{-Mt}~G(\xi,t),\label{Mmom}
\eeq
and $\chi(M,\as)$ is the BFKL function which is now known at NLO accuracy:
\footnote{Note that the normalization conventions for $\chi_0$ and $\chi_1$ 
used here are different from those used in either of refs.\cite{fl,sxap}.}
\beq
\chi(M,\as)=\as\chi_0(M)+\as^2\chi_1(M)+\dots\>.\label{chiexp}
\eeq
In eq.~(\ref{xevol}) the coupling $\as$ is fixed. 
The inclusion of running effects in the BFKL theory is a delicate
point. To next-to-leading order in $\as$ (i.e. to NLLx),
running effects can be included~\cite{sxap,runcoup} by adding to
$\chi_1$ a term proportional to the first coefficient 
$\beta_0=\smallfrac{11}{3}n_c-\smallfrac{2}{3}n_f$ of the 
$\beta$-function. Since furthermore the extra term depends on 
the definition of the gluon density, it is also necessary to 
specify the choice of factorization scheme:
here we choose the \MS\ scheme, so that the $\chi_1$ that we will 
consider in the sequel is given by~\cite{sxap}
\beq
\chi_1(M) = \smallfrac{1}{4\pi^2} n_c^2 \tilde\delta(M) 
+\smallfrac{1}{8\pi^2}\beta_0 n_c
((2\psi'(1)-\psi'(M)-\psi'(1-M))
+\smallfrac{1}{4n_c^2}\chi_0(M)^2,\label{chionedef}
\eeq
where the function $\tilde\delta$ is defined in the first of 
ref.~\cite{fl}.

\begin{figure}
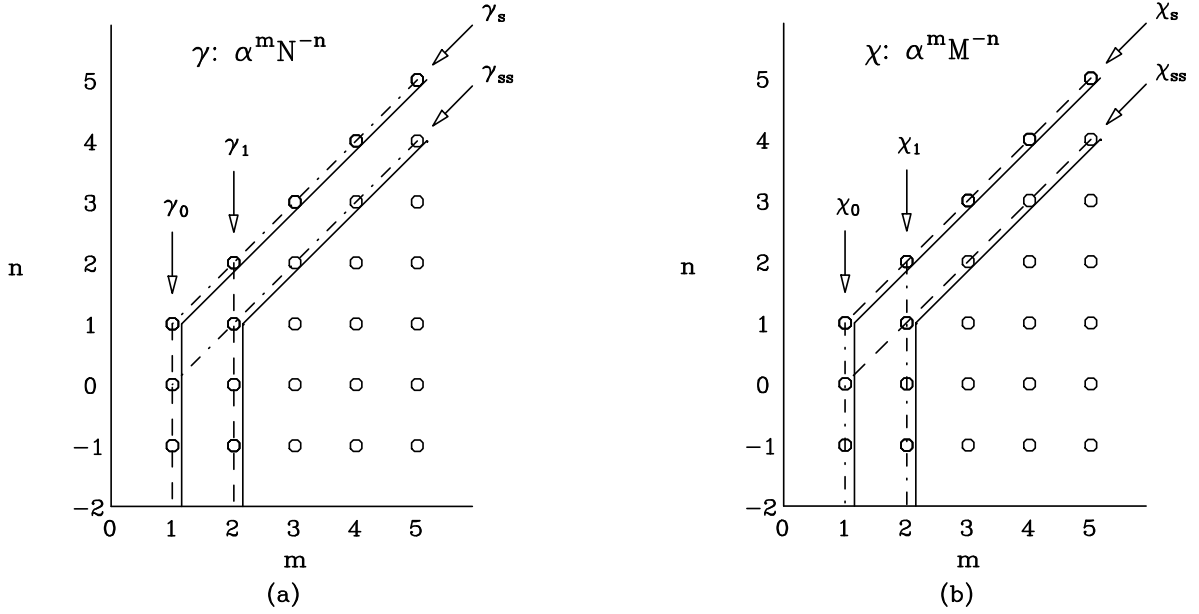

\begin{center}
\mbox{\hbox{
 \epsfig{file=fig1ax.ps,width=0.4\textwidth}}
\hskip 2truecm\hbox{\epsfig{file=fig1bx.ps,
width=0.4\textwidth}}}
\caption{\baselineskip 10pt Graphical representation of different 
expansions of (a) $\gamma$  and (b) $\chi$  in powers of 
$\as$ and $1/N$ (a) and $1/M$ (b) to order $m$ and $n$ respectively, 
and the different relations between these expansions. 
Vertical lines correspond to terms
of the same fixed order in $\as$: for example the one loop anomalous
dimension $\gamma_{0}$ contains terms with $m=1$, $n=1,0,-1,-2,\ldots$. 
Diagonal lines correspond to terms of the same order in 
$\as$ at fixed $\as\over N$ (a) or $\as\over M$ (b): for 
example $\gamma_s(\as/N)$  contains
terms with $m=n=1,2,3,\ldots$.  The sum of terms in
a vertical line of the $\gamma$ plot is related by duality 
eqs.~(\ref{dual},\ref{revdual})
to the sum of terms  in a diagonal line in the $\chi$  plot and
conversely (marked by the same line style).
The solid lines denote terms of the same order in the ``envelope''
or ``double leading'' expansion discussed in the text.
}{
}
\end{center}
\end{figure}

In the region where $Q^2$ and $1/x$ are both large the $t$ and
$\xi$ evolution equations, \ie~eqs.(\ref{tevol},\ref{xevol}),  
are simultaneously valid, and their mutual consistency requires the
validity of the ``duality" relation~\cite{jar,afp}:
\beq
\chi(\gamma(N,\as),\as)=N,\label{dual}
\eeq
and its inverse
\beq
\gamma(\chi(M,\as),\as)=M.\label{revdual}
\eeq
Using eq.~(\ref{dual}), knowledge of the expansion 
eq.~(\ref{chiexp}) of $\chi(M,\as)$ 
to LO and NLO in $\as$ at fixed $M$ 
determines the coefficients of the expansion of 
$\gamma(N,\as)$ in powers of $\as$ at fixed $\smallfrac{\as}{N}$:
\beq
\gamma(N,\as)=\gamma_s(\smallfrac{\as}{N})
+\as\gamma_{ss}(\smallfrac{\as}{N})+\ldots,\label{sxexp} 
\eeq
where $\gamma_s$ and $\gamma_{ss}$ contain respectively sums of 
all the leading and subleading singularities of $\gamma$  (see fig.~1),
\bea
\chi_{0}(\gamma_{s}(\smallfrac{\as}{N}))&=&{N\over\as},\label{dual0}\\
\gamma_{ss}(\smallfrac{\as}{N})&=&
-\frac{\chi_{1}(\gamma_{s}(\smallfrac{\as}{N}))}
{\chi'_{0}(\gamma_{s}(\smallfrac{\as}{N}))}.\label{dual1}
\eea
This corresponds to an expansion of the splitting function 
in logarithms of $x$: if for example we write
\beq
\gamma_{s}(\frac{\as}{N})=
\sum_{k=1}^{\infty}g_k^{(s)}\left(\frac{\as}{N}\right)^k
\label{ga0}
\eeq
(where $g_1^{(s)}=n_c/\pi,\>g_2^{(s)}=g_3^{(s)}=0\>,
g_4^{(s)}=2 \zeta(3) (n_c/ \pi)^4,\dots$),
then the associated splitting function
\beq
P_s(\as\xi)\equiv  \int_{c-i\infty}^{c+i\infty}\!\frac{dN}{2\pi i\as}\,
e^{N\xi}\,\gamma_s(\smallfrac{\as}{N})=
\sum_{k=1}^{\infty}\frac{g_k^{(s)}}{(k-1)!}(\as\xi)^{(k-1)},\label{Pa}
\eeq
and similarly for the subleading singularities $P_{ss}(\as\xi)$, etc.

Likewise, the inverse duality eq.~(\ref{revdual}) relates
the fixed order expansion eq.~(\ref{gamexp}) of $\gamma(N,\as)$ to 
an expansion of $\chi(M,\as)$ in powers of $\as$ with 
$\smallfrac{\as}{M}$ fixed: if 
\beq
\chi(M,\as)
=\chi_s(\smallfrac{\as}{M})+\as\chi_{ss}(\smallfrac{\as}{M})
+\ldots,\label{clqexp} 
\eeq
where now $\chi_s(\smallfrac{\as}{M})$ and 
$\chi_{ss}(\smallfrac{\as}{M})$ contain the leading 
and subleading singularities respectively of $\chi(M,\as)$, then
\bea
\gamma_{0}(\chi_{s}(\smallfrac{\as}{M}))&=&\frac{M}{\as},
\label{revdual0}\\
\chi_{ss}(\smallfrac{\as}{M})&=&
-\frac{\gamma_{1}(\chi_{s}(\smallfrac{\as}{M}))}
{\gamma'_{0}(\chi_{s}(\smallfrac{\as}{M}))}.
\label{revdual1}
\eea

In principle, since $\chi_0$ and $\chi_1$ are known, they can be 
used to construct an improvement of the splitting function which 
includes a summation of leading and subleading logarithms of $x$. 
However, as is now well known, 
the calculation~\cite{fl,cc,dd} of $\chi_{1}$ 
has shown that this procedure is confronted with serious problems.
The fixed order expansion eq.~(\ref{chiexp}) is very badly behaved: at 
relevant values of $\as$ the NLO term
completely overwhelms the LO term. 
In particular, near $M=0$, the behaviour is unstable, with $\chi_0\sim 1/M$,
$\chi_1\sim -1/M^2$. 
Also, the value of $\chi$ near the
minimum is subject to a large negative NLO correction, 
which turns the minimum into a maximum and can even reverse the sign
of $\chi$ at the minimum. 
Finally, if one considers the
resulting $\gamma_s$ and $\gamma_{ss}$ or their Mellin transforms 
$P_s(x)$ and $P_{ss}(x)$ one finds that the NLO terms
become much larger than the LO terms and negative in the region of
relevance for the HERA data~\cite{flph}.  We now discuss our
proposals to deal with all these problems.

Our first observation is that a much more stable expansion for $\chi(M)$ 
can be obtained if we make appropriate use of the additional information 
which is contained in the one and two loop anomalous dimensions 
$\gamma_{0}$ and $\gamma_{1}$. Instead of trying to improve the fixed 
order expansion eq.~(\ref{gamexp}) of $\gamma$ by all order summation 
of singularities deduced from the fixed order expansion eq.~(\ref{chiexp}) 
of $\chi$, we attempt the converse: we improve 
$\chi_{0}(M)$ by adding to it the all order summation of 
singularities $\chi_{s}$ eq.~(\ref{revdual0}) deduced 
from $\gamma_0$, $\chi_{1}(M)$ by adding to it $\chi_{ss}$ deduced 
from $\gamma_1$  eq.~(\ref{revdual1}), and so on. It can then be
seen that the instability at $M=0$ of the usual fixed order expansion 
of $\chi$ was inevitable: momentum conservation for the 
anomalous dimension, eq.~(\ref{mcons}), implies, given the 
duality relation, that the value of $\chi(M)$ at $M=0$ is fixed 
to unity, since from eq.~(\ref{dual}) we see that at $N=1$
\beq
\chi(\gamma(1,\as),\as)=\chi(0,\as)=1.\label{dualmcons}
\eeq
It follows that the fixed order expansion of $\chi$ must be poorly
behaved near $M=0$: a simple 
model of this behaviour is to think of replacing $\as/M$ with 
$\as/(M+\as)=\as/M - \as^2/M^2+...$ in order to 
satisfy the momentum conservation constraint. 
We thus propose a reorganization of the expansion of $\chi$ into 
a ``double leading'' (DL) expansion, organized in terms of ``envelopes'' 
of the contributions summarized in fig.~1b: each order contains a ``vertical''
sequence of terms of fixed order in $\as$, supplemented by a ``diagonal'' 
resummation of singular terms of the same order in $\as$ if $\as/M$ is 
considered fixed. To NLO the new expansion is thus
\bea
&\chi(M,\as)&=\left[\as\chi_{0}(M)
+\chi_{s}\left({\smallfrac{\as}{M}}\right)-
\smallfrac{n_c\as}{\pi M}\right]\nonumber\\
&&\qquad +\as\left[\as\chi_{1}(M)
+\chi_{ss}\left({\smallfrac{\as}{M}}\right)-\as\left(\smallfrac{f_2}{M^2}+
\smallfrac{f_1}{M}\right)-f_0\right]+\cdots\label{cdl}
\eea
where the LO and NLO terms are contained in the respective square brackets. 
Thus the LO term contains three contributions: $\chi_0(M)$ is the 
leading BFKL function~eq.~(\ref{chiexp}), $\chi_{s}(\as/M)$ 
eq.~(\ref{revdual0}) are resummed leading singularities 
deduced from the one loop anomalous dimension, and
$n_c\as/(\pi M)$ is subtracted  to avoid double
counting. At LO the momentum conservation 
constraint~eq.~(\ref{dualmcons}) is satisfied exactly 
because $\gamma_{0}(1)=0$ and
$[\chi_0(M)-\smallfrac{n_c}{\pi M}]\sim M^2$ near $M=0$.
At NLO there are again three types of contributions:  $\chi_1(M)$ from 
the NLO fixed order calculation (eq.~(\ref{chionedef})), the 
resummed subleading singularities $\chi_{ss}(\as/M)$ deduced 
from the two loop anomalous dimension, 
and three double counting terms, $f_0=0$, $f_1= 
-n_f(13+10n_c^2)/(36\pi^2n_c^3) $ and
$f_2=n_c^2(11+ 2 n_f/n_c^3)/(12\pi^2)$ (corresponding to those terms with
$(m,n)=(1,0),(2,1),(2,2)$ respectively in fig.~1b). 
Note that at the next-to-leading level the momentum conservation 
constraint is not exactly satisfied because the constant 
contribution to $\chi_1$ does not vanish in \MS, even though 
it is numerically very small (see fig.~2).
It could be made exactly zero by a refinement of the double 
counting subtraction but we leave further discussion of this point for 
later.

\begin{figure}[t!]
\begin{center}
    \mbox{
      \epsfig{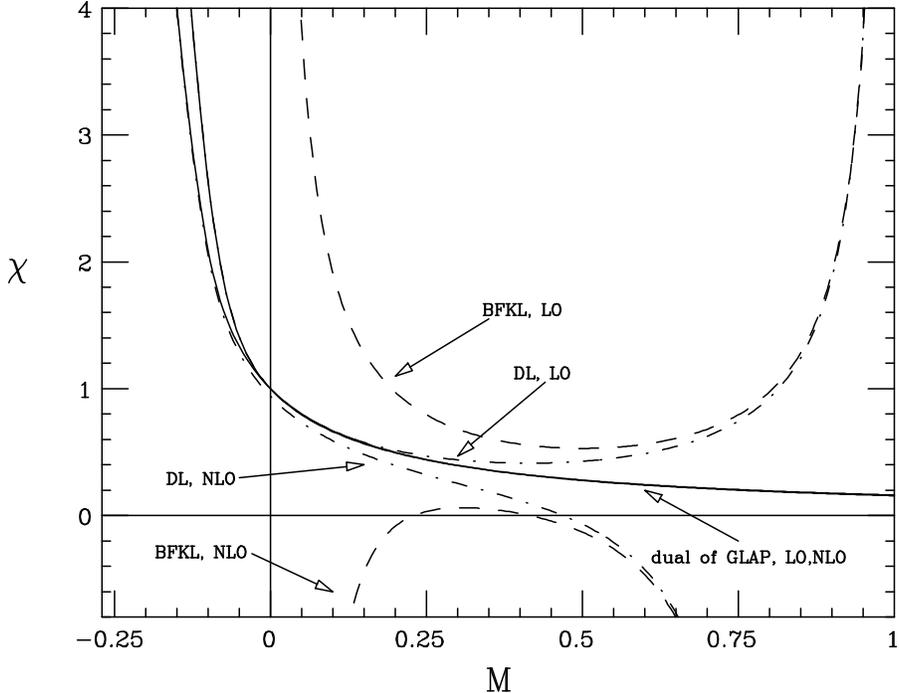}
      }\caption{\baselineskip 10pt 
      Plots of different approximations to the $\chi$ function
      discussed in the text: the BFKL leading and next-to-leading 
order functions eq.~(\ref{chiexp}), $\as\chi_0$ and $\as\chi_0+\as^2\chi_1$
(dashed); the LO and NLO dual $\as \chi_{s}$ and 
$\as \chi_{s}+ \as^2 \chi_{ss}$  of the one and two loop anomalous dimensions
(solid), and the double-leading functions at LO and NLO defined in  
eq.~(\ref{cdl}) (dotdashed). Note that the double-leading curves coincide with 
the resummed ones at small $M$, and with the fixed order ones at large $M$.
}{}
\end{center}
\end{figure}

Plots of the various LO and NLO approximations to $\chi$ are 
shown in fig.~2. In this and other plots in this paper we take
$\as =0.2$, which is a typical value in the HERA region,
and the number of active flavours $n_f=4$. 
We see that, as discussed above, the usual fixed order expansion
eq.~(\ref{chiexp}) in terms of $\chi_0$  and $\chi_1$ is 
very unstable. However, the new 
expansion eq.~(\ref{cdl}) is stable up to $M\lappeq 0.3-0.4$. 
Furthermore, in this region, $\chi$ evaluated in the double leading 
expansion (\ref{cdl}) is very close to the resummations of leading and 
subleading singularities eq.~(\ref{clqexp}) obtained by duality 
eq.~(\ref{revdual0},\ref{revdual1}) from the one and two loop 
anomalous dimensions. This shows that in this region the dominant 
contribution to $\chi$, and thus to $\gamma$ , comes from the 
resummation of logarithms of $Q^2/\mu^2$ with $Q^2\gg \mu^2$.

Beyond $M\sim 0.4$, the size of the 
contributions from collinear singular and nonsingular terms becomes 
comparable (after all here $Q^2\sim\mu^2$), but the calculation of the latter 
(from the fixed expansion eq.~(\ref{chiexp})) has become unstable due to 
the influence of the singularities at $M=1$. No complete 
and reliable description of $\chi$ seems possible without some sort of 
stabilization of these singularities. However, since they correspond 
to infrared singularities of the BFKL kernel (specifically logarithms of 
$Q^2/\mu^2$ with $Q^2\ll \mu^2$) this  would necessarily 
be model dependent. In particular, such a stabilization cannot be
easily deduced from the resummation of the $M=0$ singularities:
the original symmetry of the gluon--gluon amplitude at large $s$ is
spoiled by running coupling effects and by unknown effects from
the coefficient function through which it is related 
to the deep-inelastic structure functions, in a way which 
is very difficult to control near the photoproduction limit 
$M\to 1$. We thus prefer not to enter into this problem: rather we  
will discuss later a practical procedure to bypass it.

The results summarized in fig.~2 clearly illustrate the superiority 
of the new double leading expansion of $\chi$ over the fixed order 
expansion, and already indicate that the complete 
$\chi$ function could after all lead to only small departures from 
ordinary two loop evolution.

\bigskip\noindent
{\bf 3.}~~~Having constructed a more satisfactory expansion~eq.~(\ref{cdl}) 
of the kernel $\chi$, we now derive from it an improved form of the anomalous
dimension $\gamma$ to be used in the evolution~eq.~(\ref{tevol}), 
in order to achieve a more complete description 
of scaling violations valid both at large and small $x$. In principle, 
this can be done by using the duality relation eq.~(\ref{dual}), 
which simply gives the function $\gamma$ as the inverse of the 
function $\chi$. However, in order to derive an analytic 
expression for $\gamma(N,\as)$ which also allows us
to clarify the relation to previous attempts we
start from the naive double-leading expansion of 
$\gamma$~\cite{summ} in which 
terms are organized into ``envelopes'' of the contributions 
summarized in fig.~1a in an analogous way to the double leading 
expansion (\ref{cdl}) of $\chi$: 
\bea
&\gamma(N,\as)&=\left[\as\gamma_{0}(N)
+\gamma_{s}\left(\smallfrac{\as}{N}\right)-
\smallfrac{n_c\as}{\pi N}\right]\nonumber\\
&&\qquad +\as\left[\as\gamma_{1}(N)
+\gamma_{ss}\left(\smallfrac{\as}{N}\right)
-\as\left(\smallfrac{e_2}{N^2}+
\smallfrac{e_1}{N}\right)-e_0\right]+\cdots,\label{gdl}
\eea
where now $e_2=g_2^{(s)}=0$, $e_1=g_1^{(ss)}=
n_f n_c (5+13/(2n_c^2))/(18\pi^2)$ 
and $e_0=-(\smallfrac{11}{2}n_c^3+ n_f)/(6\pi n_c^2)$.
In this equation, the leading and subleading singularities $\gamma_{s}$ and
$\gamma_{ss}$ are obtained using duality eq.~(\ref{dual}) 
from $\chi_0$ and $\chi_1$, and summed up to give expressions which 
are exact at NLLx. These are then added to the usual one and two loop 
contributions, and the subtractions take care of the double counting 
of singular terms. 

It can be shown that the dual of 
the double leading expansion of $\chi$ eq.~(\ref{cdl}) coincides 
with this double leading expansion of $\gamma$  eq.~(\ref{gdl}) 
order by order in perturbation theory, up to terms which
are higher order in the sense of the double leading expansions. 
However, it is clear that these additional subleading terms must be
numerically important. Indeed, it is well know that at small $N$ the
anomalous dimension in the small-$x$ expansion eq.~(\ref{sxexp})
is completely dominated by $\gamma_{ss}(\as/N)$ which grows very 
large and negative, leading to completely unphysical results in the 
HERA region~\cite{flph}. It is clear that this perturbative 
instability will also be a problem in the 
double leading expansion eq.~(\ref{gdl}). On the other hand, 
we know from fig.~2 that the exact dual of $\chi$ in double 
leading expansion is stable, and 
not too far from the usual two loop result. The 
origin of this instability problem, and a suitable reorganization of the 
perturbative expansion which allows the resummation of the 
dominant part of the subleading terms have been discussed 
in ref.~\cite{sxap}. After this resummation, the resulting 
expression for $\gamma$ in double leading expansion 
will be very close to the exact dual of the corresponding expansion 
of $\chi$. 

The procedure of ref.~\cite{sxap} can be interpreted in a simple way
whenever the all-order ``true'' function $\chi(M,\as)$ 
possesses a minimum at a real value of $M$, $M_{min}$, 
with $0<M_{min}<1$ (although the final result for the
anomalous dimension will retain its validity even in the absence 
of such minimum). Using $\lambda$ to denote this minimum value of $\chi$,
\beq
\lambda\equiv\chi(M_{min},\as)=\lambda_0+\Delta \lambda, 
\qquad \lambda_0\equiv\as \chi_0(\half)= \smallfrac{4n_c}{\pi} \as \ln 2. 
\label{lamdef}
\eeq
The instability turns out to be due to the fact that higher order 
contributions to $\gamma$ must change the
asymptotic small $x$ behaviour from $x^{-\lambda_0}$ to $x^{-\lambda}$.
The starting point of the proposed procedure consists of absorbing 
the value of the correction to the value of $\chi$ at the minimum
into the leading order term in the expansion of $\chi$:
\bea
\chi(M,\as)&=&\as\chi_0(M)+ \as^2\chi_1(M)+\dots\nonumber\\
      &=&(\as\chi_0(M)+\Delta\lambda)+ \as^2\tilde\chi_1(M)+\cdots,
\label{tilsub}
\eea
where $\tilde\chi_n(M)\equiv\chi_n(M)-c_n$, with $c_n$ chosen so that
$\tilde\chi_n(M)$ no longer leads to an $O(\as^n)$ shift in the minimum.
Since the position $M_{min}$ of the all-order minimum is not known,
one must in practice expand it in powers of $\as$ around the leading order
value $M=\half$, so at higher orders the expressions for the subtraction 
constants $c_n$ can become quite complicated functions of $\chi_i$ and 
their derivatives at $M=\half$~\cite{sxap}. However at NLO 
we have simply $c_1=\chi_1(\half)$, so 
$\Delta\lambda=\as^2\chi_1(\half)+\cdots$.

A stable expansion of $\gamma$ in resummed leading and 
subleading singularities can now be obtained from the 
duality eqs.(\ref{dual0},\ref{dual1},\dots) by treating
$\chi_0+\Delta\lambda$ as the LO contribution to $\chi$, and the 
subsequent terms $\tilde\chi_i$ as perturbative corrections to it.
Of course, since the reorganization eq.~(\ref{tilsub}) amounts 
to a reshuffling of perturbative orders, to any finite order the
anomalous dimension obtained in this way will be equal to the 
old one up to formally subleading  corrections.
Explicitly, we find in place of the previous expansion in sums of 
singularities eq.~(\ref{sxexp}) the resummed expansion
\beq
\gamma(N,\as)=\gamma_s\left(\smallfrac{\as}{N-\Delta \lambda}\right)+
\as\tilde\gamma_{ss}\left(\smallfrac{\as}{N-\Delta \lambda}\right)
+\dots,\label{gamimpr} 
\eeq
where 
\beq
\tilde\gamma_{ss}\left(\smallfrac{\as}{N-\Delta \lambda}\right)
\equiv \gamma_{ss}\left(\smallfrac{\as}{N-\Delta \lambda}\right)
+{\chi_1(\half)\over\chi^\prime_0\left(\gamma_s
\left(\smallfrac{\as}{N-\Delta \lambda}\right)\right)}.
\eeq
In terms of splitting functions this resummed expansion is simply 
\bea
xP(x,\as)&=&\as e^{\xi\Delta\lambda}
[P_{s}(\as\xi)+\as \tilde P_{ss}(\as\xi)+\dots]\nonumber\\
&=&\as e^{\xi\Delta\lambda}
[P_{s}(\as\xi)+\as P_{ss}(\as\xi)
-\xi\Delta\lambda P_{s}(\as\xi)+\dots].\label{Pimpr}
\eea
The expansion is now stable~\cite{sxap}, 
in the sense that it may be shown that 
$\tilde P_{ss}(\as\xi)/P_{s}(\as\xi)$ remains bounded as 
$\xi\to\infty$: subleading corrections will then be small 
provided only that $\alpha_s$ is sufficiently small.
This result may be shown to be true to all orders in perturbation 
theory, using an inductive argument. 

We can thus replace the unresummed singularities $\gamma_s$ 
and $\gamma_{ss}$ in eq.~(\ref{gdl})  with
the resummed singularities eq.~(\ref{gamimpr})
to obtain a double leading expansion with stable small $x$ behaviour:
\bea 
\gamma(N,\as)&=& \left[\as\gamma_{0}(N)+
\gamma_{s}(\smallfrac{\as}{N-\Delta \lambda})
-\as\smallfrac{n_c}{\pi N}\right]\nonumber \\ &&
+\as\left[\as\gamma_{1}(N)
+\tilde\gamma_{ss}(\smallfrac{\as}{N-\Delta \lambda})
-\as(\smallfrac{e_2}{N^2}+\smallfrac{e_1}{N}) - e_0\right]+\cdots.
\label{til}
\eea
Momentum conservation is violated by the resummation  
because $\gamma_{s}$ and $\gamma_{ss}$ and the subtraction terms 
do not vanish at $N=1$. It can be restored by 
simply adding to the constant $e_0$ a further series of constant terms
beginning at $O(\as^2)$: these are all formally subleading in 
the double leading expansion. This constant shift in $\gamma$ is
precisely analogous to the shift made on $\chi$ in
eq.~(\ref{lamdef}) which generated the resummation. 

It is important to recognize that there is inevitably an 
ambiguity in the double counting subtraction terms 
in eq.~(\ref{til}). For example, at the leading order 
of the double leading expansion instead of subtracting  
$ n_c \as\over\pi N$ we could have subtracted 
$n_c\as\over \pi(N-\Delta\lambda)$, since this
differs only by formally subleading terms: $\Delta\lambda=O(\as^2)$, so
\beq
{\as\over N}={\as\over
N-\Delta \lambda}\left(1-{\Delta \lambda\over N
-\Delta\lambda}+\dots\right).\label{tiltilexp}
\eeq
Following the same type of subtraction at NLO, the resummed double leading 
anomalous dimension may thus be written as  
\bea 
\gamma(N,\as)&=& \left[\as\gamma_{0}(N)+
\gamma_{s}(\smallfrac{\as}{N-\Delta \lambda})
-\as\smallfrac{n_c}{\pi (N-\Delta \lambda)}\right]\nonumber \\ &&
+\as\left[\as\gamma_{1}(N)
+\tilde\gamma_{ss}(\smallfrac{\as}{N-\Delta \lambda})
+\smallfrac{n_c\Delta\lambda}{\pi(N-\Delta \lambda)^2}
-\as(\smallfrac{e_2}{(N-\Delta \lambda)^2}
+\smallfrac{e_1}{N-\Delta \lambda}) - e_0\right]+\cdots.
\label{tiltil}
\eea
The extra term at NLO comes from the first correction in 
eq.~(\ref{tiltilexp}), which is of order $\as^3\over
N^2$, and thus a subleading singularity. 
The characteristic feature of this alternative resummation 
is that the fixed order anomalous dimensions $\gamma_{0}$, $\gamma_{1}$ 
are preserved in their entirety, including the position of their 
singularities. As with the previous expansion eq.~(\ref{til}) 
momentum conservation may be imposed by adding to $e_0$ a 
series of terms constant in $N$ and starting at $O(\as^2)$.

This completes our procedure of inclusion of the most important part
of the subleading corrections, as we shall see shortly by a direct comparison
of the resummed expansions eq.~(\ref{til}) and eq.~(\ref{tiltil})
with the exact dual of $\chi$ evaluated according to eq.~(\ref{cdl}).
In the sequel we will discuss the phenomenology based on the two 
resummed expansions eq.~(\ref{til}) and eq.~(\ref{tiltil}) on an 
equal footing, taking the spread of the results as an indication of
the residual ambiguity due to subleading terms. Although formally the 
differences between the two expansions are subleading, we will find that 
in practice they may be quite substantial, because 
$\Delta\lambda$ may be large.

\begin{figure}[t!]
\begin{center}
    \mbox{
      \epsfig{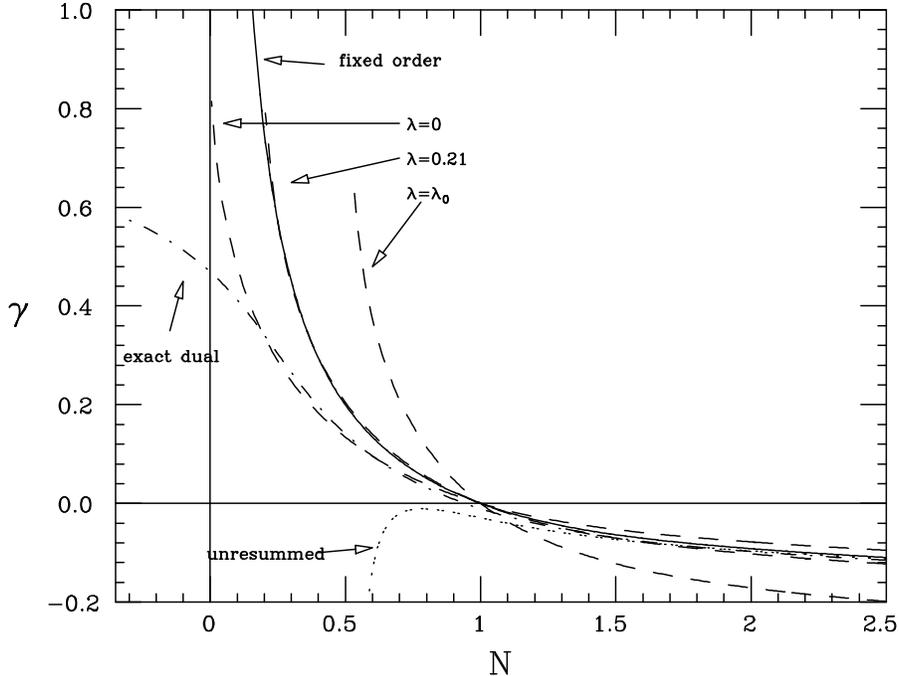}
      }
\caption{\baselineskip 10pt 
Comparison of the anomalous dimension $\gamma$ evaluated at NLO in the 
resummed expansion eq.~(\ref{til}) for three different values of $\lambda$  
(dashed) with the usual fixed order perturbative anomalous dimension 
(also at NLO) eq.~(\ref{gamexp}) (dotted) and that obtained by 
exact duality from $\chi$ at NLO in the expansion eq.~(\ref{cdl}) as displayed
in fig.~2 (solid).  The unresummed $\gamma$ eq.~(\ref{gdl})
is also shown at NLO. Notice that the $\lambda=0.21$ curve is very close to
the two loop anomalous dimension down to the branch point at $N=\lambda$. }
\vskip -.8truecm
{}
\end{center}
\end{figure}

\bigskip\noindent
{\bf 4.}~~~So far we have constructed resummations of the anomalous 
dimension and splitting function which satisfy the elementary 
requirements of perturbative stability and momentum conservation. 
This construction
relies necessarily on the value $\lambda$ of $\chi$ near its minimum, 
since it is this which determines the small $x$ behaviour of 
successive approximations to the splitting function. In order to obtain a 
formulation that can be of practical use for actual phenomenology, 
we will need however to improve the description of $\chi(M)$ in 
the ``central region'' near its minimum $M_{min}$, since 
as we already observed, we cannot reliably
determine the position and value of the minimum of $\chi$ without a
stabilization of the $M=1$ singularity. Indeed, we can see from fig.~2
that in the central region $\chi$ evaluated in the double 
leading expansion is dominated by the presumably
unphysical $M=1$ poles of $\chi$, and at NLO this means that it 
actually has no minimum, becoming rapidly negative. However, 
one can use the value $\lambda$ of the true $\chi$ at the 
minimum as a useful parameter for an effective
description of the $\chi$ function around $M=1/2$.
Indeed, $\Delta \lambda$ as estimated from its next-to-leading order
value $\as^2 \chi_1(1/2)$ turns out to be of the same order as
$\lambda_0$ for plausible values of $\as$, a feature which can
be also directly seen from fig.~2. This supports the idea that $\lambda$ and
$\Delta \lambda$ are not truly perturbative quantities: in general we
expect that the overall shift of the minimum will still be of the
order of $\lambda_0$ and negative. It is this order transmutation that
makes the impact of the resummations eq.~(\ref{til},\ref{tiltil}), and 
the differences between them, quite substantial.

\begin{figure}[t!]
\vskip-.1truecm
\begin{center}
    \mbox{
      \epsfig{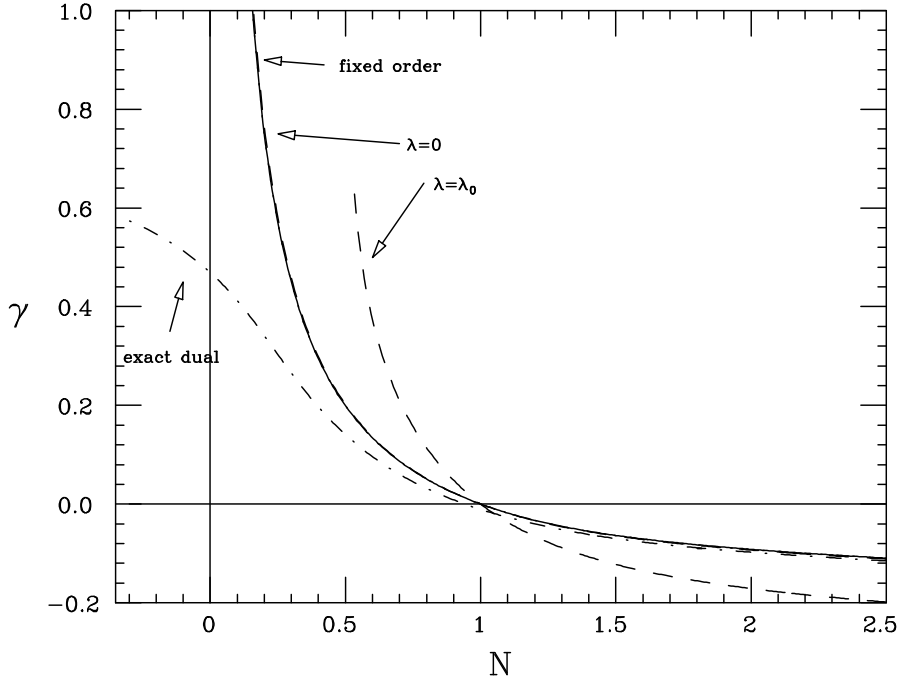}
      }
\caption{
\baselineskip 10pt 
Comparison of the anomalous dimension $\gamma$ evaluated at NLO in the 
resummed expansion eq.~(\ref{tiltil})
for two different values of $\lambda$  (dashed) with 
the usual fixed order perturbative anomalous dimension (also at NLO) 
eq.~(\ref{gamexp}) (solid) and that obtained by exact duality from  
$\chi$ at NLO as in fig.~3 (dotdash). 
Notice that the $\lambda=0$ curve is virtually 
indistinguishable from  the fixed order anomalous dimension 
for all values of $N$.}
\vskip -.8truecm
{}
\end{center}
\end{figure}

In fig.~3  and fig.~4 we display the results for the resummed
anomalous dimensions in the two different expansions eq.~(\ref{til}) and 
eq.~(\ref{tiltil}) respectively, each computed at next-to-leading order. 
In both figures we show for comparison the fixed order anomalous dimension
$\as\gamma_{0}(N)+\as^2\gamma_{1}(N)$ 
eq.~(\ref{gamexp}). Also for comparison, we show the exact dual 
of $\chi$ computed at NLO in the double leading expansion 
eq.~(\ref{cdl}), obtained from eq.~(\ref{dual}) by
exact numerical inversion. This curve is thus simply the 
inverse of the corresponding curve already shown in fig.~2. 

In fig.~3 we show the anomalous dimension computed at NLO using the 
resummation eq.~(\ref{til}), for $\lambda=\lambda_0$ and $\lambda=0$. 
The first value  corresponds to  the LO
approximation to $\lambda$, while  the second value is close to the
NLO approximation when $\as$ is in the region
$\as\sim 0.1 - 0.2$. 
We might expect the value of $\lambda$ as determined by the
actual all-order minimum of $\chi$  to lie within this range.   
Note that,  in general, the resummed anomalous dimension has a cut
starting at $N=\lambda$, which corresponds to the $x^{-\lambda}$ power
rise; for this reason  our plots  stop at this value of $N$.  
The $\lambda=0$ curve, corresponding to the next-to-leading order
approximation to $\lambda$, is seen to be very close to the exact dual
of $\chi$ at NLO in the expansion eq.~(\ref{cdl}), as already anticipated. 
This is to be contrasted with the corresponding unresummed anomalous dimension
eq.~(\ref{gdl}), which is also  displayed in
fig.~3, and is characterized by the rapid fall at small $N$ discussed
already in ref.~\cite{flph,sxap}. This comparison demonstrates that indeed
the perturbative reorganization eliminates this pathological steep
decrease. The resummed curve with $\lambda=0$ and the exact
dual of $\chi$ become rather different for small $N\lsim
0.2$. However, this is precisely the range of $N$ which corresponds
to the central region of $M$ where we cannot trust the next-to-leading
order determination of $\chi$. 
Finally, we show that we can choose a value of $\lambda\simeq 0.21$ such
that the resummed anomalous dimension closely reproduces the two loop
result down to the branch point at $N=\lambda$. This shows
that the absence of visible deviations from the usual two loop evolution
can be accommodated by the resummed anomalous dimension. However this
is not necessarily the best option phenomenologically: perhaps 
the data could be better fitted
by a different value of $\lambda$ if a suitable modification of the input
parton distributions is introduced. It is nevertheless clear that large
values of $\lambda$ such as $\lambda \approx\lambda_0$ can be easily
excluded within the framework of this resummation, since they would
lead to sizeable deviations from the standard two loop scaling
violations in the medium and large $x$ region.

The splitting functions corresponding to the anomalous dimensions of
fig.~3 are displayed in fig.~5. The
basic qualitative features are of course preserved: in particular,
the curves with small values of  $\lambda=0$ and $\lambda=0.21$ 
are closest to the two loop
result. However, on a more quantitative level, it is clear that
anomalous dimensions which coincide in a certain range of $N$, 
but differ in other regions (such as very small $N$)
may lead to splitting functions which differ over a considerable region 
in $x$. In particular, the 
$\lambda=0.21$ curve displays the predicted $x^{-\lambda}$ growth
at sufficiently large $\xi$ ($x\lsim 10^{-4}$). The dip seen
in the figure for intermediate values of $\xi$ is necessary in order
to compensate this growth in such a way that 
the moments for moderate values of $N$ remain unchanged. Note that
the $x^{-\lambda}$ behaviour of the splitting function at small $x$
is corrected by logs~\cite{sxap}:  
$P_s\toinf\xi \xi^{-3/2} x^{-\lambda}$. If $\lambda=0$ this
logarithmic drop provides the dominant large $\xi$ behaviour which
appears in the figure.

If the anomalous dimensions are instead resummed as in eq.~(\ref{tiltil}),
the results are as shown in fig.~4, again for the two very different  
values of $\lambda$, $\lambda=0$ and $\lambda=\lambda_0$. 
When $\lambda=0$ the resummed anomalous
dimension is now essentially indistinguishable from
the two loop result. This is due to the fact that the simple
poles at $N=0$ which are now retained in $\gamma_{0}$ and $\gamma_{1}$   
provide the dominant small $N$ behaviour. The branch point at 
$N=\lambda$ in $\gamma_s$ and $\gamma_{ss}$ is then relatively subdominant. 
This remains of course true for all $\lambda\le 0$,
and in practice also for small values of $\lambda$ such as
$\lambda\lsim 0.1$. When instead $\lambda=\lambda_0$ 
the result does not differ appreciably from the resummed 
anomalous dimension shown in fig.~3, since now the dominant
small $N$ behaviour is given by the branch point at $N=\lambda_0$,
which is not affected by changes in the double counting prescription.
Summarizing, the peculiar feature of the resummation eq.~(\ref{tiltil})
is that it leads to results which are extremely close to
usual two loops for any value of $\lambda\le 0$, without the need for a
fine-tuning of $\lambda$.
\begin{figure}
\begin{center}
    \mbox{
      \epsfig{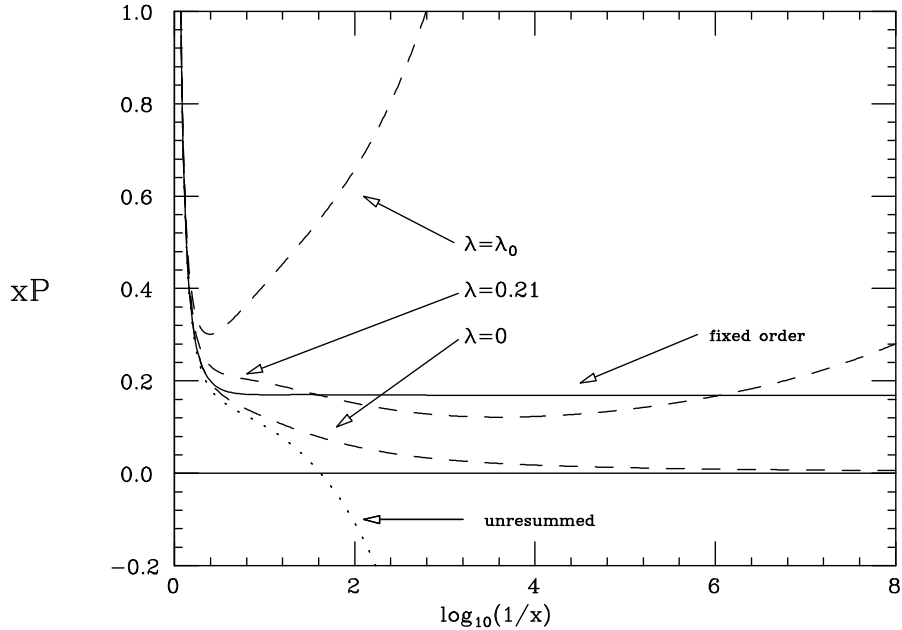}
      }
\caption{
\baselineskip 10pt 
The splitting functions corresponding to the anomalous dimensions of fig.~3.}
\vskip -.8truecm
{}
\end{center}
\end{figure}

Finally, in fig.~6 we display the splitting functions obtained from
the resummed anomalous dimensions of fig.~4. The
$\lambda=\lambda_0$ case is, as expected, very close to the corresponding
curve in fig.~5. However the $\lambda=0$ curve is now in 
significantly better agreement with
the two loop result than any of the resummed splitting functions 
of fig.~5, even that computed with the optimized value
$\lambda=0.21$. Moreover, this agreement now holds in the entire 
range of $\xi$. This is due to the fact that the corresponding 
anomalous dimension is now very close to the fixed order one 
for all $N>0$, and not only for $N>\lambda=0.21$. This demonstrates 
explicitly that one cannot exclude the possibility that the known small $x$
corrections to splitting functions resum to a result which is essentially
indistinguishable from the two-loop one. This however
is only possible if $\lambda\lsim 0$. 
\begin{figure}
\begin{center}
    \mbox{
      \epsfig{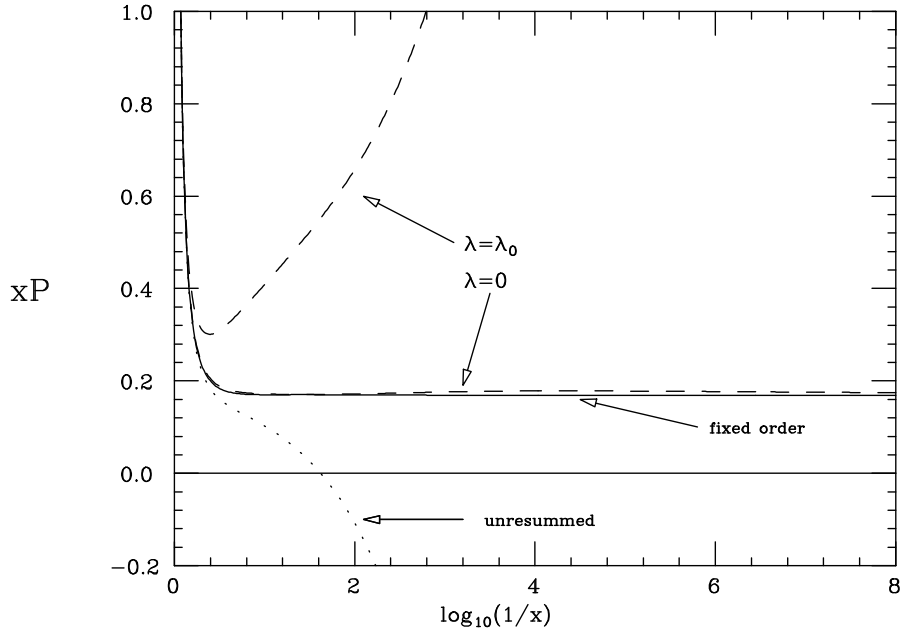}
      }
\caption{
\baselineskip 10pt 
The splitting functions corresponding to the anomalous dimensions of
fig.~4.}
\vskip -.8truecm
{

}
\end{center}
\end{figure}

To summarise, we find that the known success of perturbative evolution,
and in particular double asymptotic scaling at HERA can be
accommodated within two distinct possibilities, both of which are
compatible with our current knowledge of anomalous dimensions at small
$x$, and in particular with the inclusion of corrections derived from
the BFKL equation to usual perturbative evolution. 
One possibility, embodied by the resummed 
anomalous dimension eq.~(\ref{tiltil}) with $\lambda\lsim 0$, is that double
scaling remains a very good approximation to perturbative evolution
even if the $x \to 0$ limit is taken at finite $Q^2$. The other option,
corresponding to the resummation eq.~(\ref{til}) with a small value of
$\lambda$, is that double scaling is a 
good approximation in a wide region at
small $x$, including the HERA region, but eventually substantial 
deviations from it will show up at sufficiently small $x$. 
In the latter case, the best-fit parton distributions might be 
significantly differ from those determined at two loops even at the
edge of the HERA kinematic region. Both resummations are however 
fully compatible with a smooth matching to Regge theory in 
the low $Q^2$ region~\cite{landshoff}.

\bigskip\noindent
{\bf 5.}~~~In conclusion, we have presented a procedure for the systematic
improvement of the splitting functions at small $x$ which overcomes
the difficulties of a straightforward implementation of the BFKL
approach. The basic ingredients of our approach are the
following. First, we achieve a stabilization of the perturbative 
expansion of the function $\chi$ near $M=0$ through
the resummation of all the LO and NLO collinear singularities 
derived from the known one-- and two--loop anomalous dimensions. 
The resulting $\chi$ function is regular at $M=0$, and in fact, to a
good accuracy, satisfies the requirement imposed by momentum
conservation via duality. Then, we acknowledge that without a
similar stabilization of the $M=1$ singularity it is not possible to
obtain an reliable determination of $\chi$ in the central region
$M\sim 1/2$. However, we do not have an equally model--independent
prescription to achieve this stabilization at $M=1$. 
Nevertheless, the behaviour of $\chi$ in the central region 
can be effectively parameterized  in terms of a single parameter
$\lambda$ which fixes the asymptotic small $x$ behaviour of the
singlet parton distribution. 
This enables us to arrive at an analytic expression for the
improved splitting function, which is valid both at small and large
$x$ and  is free of perturbative instabilities. 

This formulation 
can be directly confronted with the data, which
ultimately will provide a determination of $\lambda$ along with
$\as$ and the input parton densities. The well known agreement of the
small $x$ data with the usual $Q^2$ evolution equations 
suggests that the optimal value of $\lambda$ will turn out to be
small, and possibly even negative for the relevant value of
$\alpha_s$. 
Such a value 
of $\lambda$ is theoretically attractive,
because it corresponds to a structure function whose leading-twist
component does not grow as a power of $x$ in the Regge limit: it would 
thus be compatible with unitarity constraints, and with an 
extension of the region of applicability of perturbation theory 
towards this limit.

Several alternative approaches to deal with the same problem through 
the resummation of various classes of formally subleading contributions
have been recently presented in the literature. Specific proposals 
are based on making a particular choice of the renormalization 
scale~\cite{bfklp}, or on a different
identification of the large logs which are resummed by the $\xi$ evolution
equation~(\ref{xevol}), either by a function of $\xi$
itself~\cite{schmidt}, 
or by a function of $Q^2$~\cite{salam,ciaf}, or both~\cite{for}. The main 
shortcoming of these approaches is their model dependence.  
For instance, in ref.~\cite{ciaf} the  value of $\lambda$ is
calculated, and $\chi$ is supposedly determined for all 
$0\le M\le 1$. This however requires the introduction of a 
symmetrization of $\chi$, 
which we  consider to be strongly model dependent: indeed, 
in ref.~\cite{ciaf} it is recognized that their value 
of $\lambda$  only signals the limit of applicability of their computation.
We contrast this situation with the approach to resummation presented here,
which makes maximal use of all the available model-independent 
information, with a realistic parameterization of the remaining 
uncertainties. We expect further progress to be possible 
only on the basis of either genuinely nonperturbative input, or through a 
substantial extension of the standard perturbative domain.

{\bf Acknowledgement:}  We thank S.~Catani, J.~R.~Cudell, P.~V.~Landshoff,
G.~Ridolfi and  G.~Salam for interesting discussions. This work was
supported in part by a PPARC Visiting Fellowship, and 
EU TMR  contract FMRX-CT98-0194 (DG 12 - MIHT). 
\vfill\eject


\begin{thebibliography}{99}

\bibitem{glap} V.N.~Gribov and L.N.~Lipatov, \SJNP\vyp{15}{1972}{438};\\
L.N.~Lipatov, \SJNP\vyp{20}{1975}{95};\\    
G.~Altarelli and G.~Parisi, \NP\vyp{B126}{1977}{298};\\
see also Y.L.~Dokshitzer, {it Sov.~Phys.~JETP~}\vyp{46}{1977}{691}.
\bibitem{nlo} G.~Curci, W.~Furma\'nski and R.~Petronzio, 
\NP\vyp{B175}{1980}{27};\\ E.G.~Floratos, C.~Kounnas and
R.~Lacaze, \NP\vyp{B192}{1981}{417}.
\bibitem{nnlo} S.A.~Larin, T.~van~Ritbergen, J.A.M.~Vermaseren,
\NP\vyp{B427}{1994}{41};\\  S.A.~Larin \etal, \NP\vyp{B492}{1997}{338}.
\bibitem{kis} See {\it e.g.}  R.K.~Ellis, W.J.~Stirling and
B.R.~Webber, ``QCD and Collider Physics'' (C.U.P., Cambridge 1996).
\bibitem{klein} See {\it e.g.} M.~Klein, Proceedings of the 
Lepton-Photon Symposium (Stanford, 1999),\hfil\\ {\tt 
http://www-sldnt.slac.stanford.edu/lp99/pdf/54.pdf}
\bibitem{DGPTWZ}A.~De~R\'ujula {\it et al.}, \PR\vyp{10}{1974}{1649}.
\bibitem{das} R.D.~Ball and S.~Forte, \PL\vyp{B335}{1994}{77}; 
\vyp{B336}{1994}{77};\\ {\it Acta~Phys.~Polon.~}\vyp{B26}{1995}{2097}.
\bibitem{botje} See {\it e.g.} M.~Botje (ZEUS Coll.), 
{\tt hep-ph/9905518};\\
V.~Barone, C.~Pascaud and F.~Zomer, {\tt hep-ph/9907512}.
\bibitem{bfklfits}
R.D.~Ball and S.~Forte, {\tt hep-ph/9607291};\\
I.~Bojak and M.~Ernst, \PL\vyp{B397}{1997}{296}; \NP\vyp{B508}{1997}{731};\\
J.~Bl\"umlein  and A.~Vogt, \PR\vyp{D58}{1998}{014020}.
\bibitem{bfkl} L.N.~Lipatov, \SJNP\vyp{23}{1976}{338};\\
 V.S.~Fadin, E.A.~Kuraev and L.N.~Lipatov, \PL\vyp{60B}{1975}{50};
 {\it Sov. Phys. JETP~}\vyp{44}{1976}{443};\vyp{45}{1977}{199};\\
          Y.Y.~Balitski and L.N.Lipatov, \SJNP\vyp{28}{1978}{822}.
\bibitem{jar}  T.~Jaroszewicz, \PL\vyp{B116}{1982}{291}.
\bibitem{ktfac}
S.~Catani, F.~Fiorani and G.~Marchesini, \PL\vyp{B336}{1990}{18};\\
S.~Catani et al.,  \NP\vyp{B361}{1991}{645};\\
S.~Catani and F.~Hautmann, \PL\vyp{B315}{1993}{157}; 
\NP\vyp{B427}{1994}{475}.
\bibitem{fl} V.S.~Fadin and L.N.~Lipatov, \PL\vyp{B429}{1998}{127};\\
V.S.~Fadin et al, \PL\vyp{B359}{1995}{181}; \PL\vyp{B387}{1996}{593};\\
\NP\vyp{B406}{1993}{259}; \PR\vyp{D50}{1994}{5893}; 
\PL\vyp{B389}{1996}{737};\\ \NP\vyp{B477}{1996}{767}; 
\PL\vyp{B415}{1997}{97}; \PL\vyp{B422}{1998}{287}.
\bibitem{cc} G.~Camici and M.~Ciafaloni, 
\PL\vyp{B412}{1997}{396}; \PL\vyp{B430}{1998}{349}.
\bibitem{dd} V.~del~Duca, \PR\vyp{D54}{1996}{989};
\PR\vyp{D54}{1996}{4474}; 
V.~del~Duca and C.R.~Schmidt, \PR\vyp{D57}{1998}{4069};\\  
Z.~Bern, V.~del~Duca and C.R.~Schmidt, \PL\vyp{B445}{1998}{168}.
\bibitem{flph}R.D.~Ball  and S.~Forte, {\tt hep-ph/9805315};\\
J. Bl\"umlein et al., {\tt hep-ph/9806368}.
\bibitem{salam} G.~Salam, \JHEP\vyp{9807}{1998}{19}.
\bibitem{summ} R.D.~Ball and S.~Forte, \PL\vyp{B351}{1995}{313};\\ 
R.K.~Ellis, F.~Hautmann and B.R.~Webber, \PL\vyp{B348}{1995}{582}.
\bibitem{sxap} R.D.~Ball and S.~Forte, {\tt hep-ph/9906222}
\bibitem{runcoup}G.~Camici and M.~Ciafaloni, 
\NP\vyp{B496}{1997}{305}.
\bibitem{afp} R.D.~Ball and S.~Forte, \PL\vyp{B405}{1997}{317}.
\bibitem{landshoff}  J.R. Cudell, A. Donnachie and
 P.V.~Landshoff, \PL\vyp{B448}{1999}{281}.
\bibitem{bfklp} S.J.~Brodsky \etal,  {\it JETP~Lett.~}\vyp{70}{1999}{155};\\ 
R.S.~Thorne, \PR\vyp{D60}{1999}{054031}.
\bibitem{schmidt} C.R.~Schmidt, \PR\vyp{D60}{1999}{074003}.
\bibitem{ciaf} 
M.~Ciafaloni and D.~Colferai, \PL\vyp{B452}{1999}{372};\\
M.~Ciafaloni, G.~Salam and  D.~Colferai, {\tt hep-ph/9905566}. 
\bibitem{for} J.R.~Forshaw, D.A.~Ross and A.~Sabio~Vera, 
\PL\vyp{B455}{1999}{273}.
	



\end{thebibliography}
\end{document}